\documentclass[aps,prl,preprint, groupedaddress,showpacs]{revtex4}
\usepackage[cp1251]{inputenc}
\usepackage[english]{babel}
\usepackage{epsfig}
\usepackage{amssymb}
\usepackage{amsmath}
 0
\begin{document}


 \title{Acoustic radiation force and streaming induced by non-periodic ultrasound. }
\author{  Anna Perelomova  }
\affiliation{  Gdansk University of Technology, \\
ul Narutowicza 11/12, 80-952 Gdansk, Poland, \small \\
anpe@mifgate.pg.gda.pl \\  }
\date{\today}

\newpage

\large
 \begin{abstract}
 Starting from the linear flow of homogeneous fluid, five modes are defined as
 eigenvectors of the basic system of conservation laws.
 Quasi-plane geometry is considered. Projectors that separate
 overall perturbation of the fluid into specific modes are
 calculated and applied to nonlinear flow. Dynamic equations for
 the interacting modes are obtained in the frames of the method. A
 particular case of streaming caused by acoustic pulse is
 considered, illustrations on temporal behavior of streaming velocity and streamlines
 are presented.
\end{abstract}
\maketitle                                 
\newpage

\section{Introduction}

   Streaming is known as a bulk vortical movement of the fluid
following an intense acoustic wave. Streaming appears in closed
and open vessels in two- or three-dimensional geometry and is
observed in viscous fluids only. Though vortical movement may
exist in the linear flow, the reason of streaming are the
nonlinear losses of momentum of acoustic field. A reader is
referred to the comprehensive reviews on this subject \cite{Z},
\cite{RS}, \cite{N}.
  Most recent investigations both theoretical and experimental
relate to steady (quasi-steady) streaming \cite{TT}. It is
well-known that streaming is a slow process in comparison to the
ultrasound and may be separated from the originating sound by
time-averaging over integer number of periods of sound. The
averaging is a starting point of the modern theory \cite{RS},
\cite{N}. A driving force of streaming is quadratic and therefore
does not vanish during the temporal averaging.
  Some inconsistencies of the approach based on temporal averaging
are obvious. It was first pointed out by Rudenko and Soluyan that
studying acoustic streaming while assuming incompressibility was
inadequate \cite{RS}. The effect of compressibility of the fluid
has not been discussed in depth. Recently, the importance of
compressibility has been demonstrated for flow over plane rigid
boundary \cite{Qi}. It is more evident in gases and leads to
larger streaming velocities.

We may add that incompressibility neglects not only acoustic waves
but also another slow compound of the overall flow, namely the
entropy modal field that appears in one-dimensional geometry of
flow even. Entropy mode appears due to losses of acoustic energy
and therefore has a different nature than streaming. Velocity of
this field is small in comparison to streaming \cite{MO}, the main
feature of entropy field is slow isobaric growth of temperature
that leads to the new background of acoustic wave propagation.
This effect called by acoustic wave self-action is well-studied
\cite{BZZ}. Excluding the entropy mode needs a neglecting of
equation of energy balance that reduces an initial system of
conservation laws and allows to consider flows with unperturbed
density. At last, an excluding of the energy balance does not
allow to account heat conductivity though it has been proved that
the effect of the heat conductivity could not be discarded in a
study of temperature variation associated with the streaming
\cite{MG}. Actually, this approximation is well understood in a
typical liquid like water but should be revised for other liquids
\cite{KKKB}.

  The dynamic equations are derived by proper combining the conservation
laws in differential form. These calculations may be considered as
a certain type of projecting at the chosen type of motion. We
propose to determine modes and correspondent matrix projectors
from the very beginning at the level of linearized system of
conservation laws. It seems reasonable to start from the full
system of conservation laws including the energy balance equation
and the continuity equation to avoid possible inconsistencies.

  Though ultrasound of many periods is most efficient source of
streaming, any other acoustic source in thermoviscous flow gives
rise to other modal fields. The modern theory fails to treat
non-periodic sound. That is one more reason to find new methods
(analytical more desired) in fluid dynamics and particulary in the
theory of streaming. The idea to decompose the linear flow into
specific modes is not novel and has been exploited for a long
time, see the paper by Chu, and Kovasznay \cite{CK} and referred
there papers, where homogeneous background with sources of heat
and mass are considered. This paper is almost the only one in
which interaction between modes is introduced locally (see also
report by Kaner, Rudenko, Khokhlov  \cite{KRKh}, where a first
step has been made in a manner to introduce interaction of
opposite acoustic waves via connection equations). The account of
nonlinear interactions by Chu, and Kovasznay \cite{CK} was
performed by the simple perturbation theory - by decomposition of
each variable in series of small parameters of nonlinearity and
dissipation.

The concrete ideas of doing it automatically by using of
projectors in the wide variety of problems was realized for flow
over inhomogeneous media like bubbly liquid  \cite{P1}, flows
affected by external forces including the gravitational one which
changes the background density and pressure  \cite{P2}, both one-
and multi-dimensional problems \cite{P3}). The principal advance
is an expansion of the ideas into area of nonlinear flow: to get
nonlinear coupled evolution equations for interacting modes and to
solve the system approximately. The choice of the subspaces by the
projecting (fixed links between variables) is the principal point
of evolution equations derivation.  It results in the expansion of
projection of the evolution operator as a function of a small
parameter , that is equivalent to the nonsingular perturbation
theory \cite{L}. Any weak interactions of modes in fluid dynamics
may be considered. Streaming is a particular case of the
interaction of modes when the acoustic mode is dominant and
inverse influence of the growing vortical end entropy modes
neglected.

\bigskip

\section{ Modes and projectors of the quasi-plane thermoviscous flow.
 Dynamic equations of linear and nonlinear flows.}

The mass, momentum and energy conservation laws for the
thermoviscous flow are the starting point:
\begin{equation}\label{1}
\begin{array}{c}
 \frac{\partial\rho}{\partial t}+\vec{\nabla}(\rho\vec{v})=0  \\
  \rho\left[  \frac{\partial\vec{v} }{\partial t} +(\vec{v}
\vec{\nabla} )\vec{v} \right]  =-\vec{\nabla} p+\eta\Delta\vec{v}
+\left( \varsigma +\frac{\eta}{3} \right)  \vec{\nabla}
(\vec{\nabla} \vec{v} ) \\
 \rho\left[  \frac{\partial e}{\partial
t}+(\vec{v}\vec{\nabla})e\right] +p\vec{\nabla}\vec{v}-\chi\Delta
T= \varsigma\left(  \vec{\nabla}\vec
{v}\right)  ^{2}+\frac{\eta}{2}\left(  \frac{\partial v_{i}}{\partial x_{k}%
}+\frac{\partial v_{k}}{\partial
x_{i}}-\frac{2}{3}\delta_{ik}\frac{\partial v_{l}}{\partial
x_{l}}\right)  ^{2}.
\end{array}
\end{equation}

In this system $p$,$\vec{v}$,$\rho$,$e,T$ denote pressure,
velocity, density, internal energy per unit mass and temperature
relatively, $\varsigma,$ $\eta,$ $\chi$ are bulk, shear
viscosities and thermal conductivity (all supposed to be
constants), $x_{i}$ - space coordinates, $t$-time. A system
(\ref{1}) should be completed by two thermodynamic relations for
equilibrium processes in fluid, $e(p,\rho),T(p,\rho)$ . For
simplicity, an ideal gas will be considered treated by the
following relations:
\begin{equation}\label{2}
e=\frac{p}{\rho(\gamma-1)},\hspace*{3mm}T=\frac{p}{\rho(\gamma-1)C_V}
\end{equation}
with $\gamma=C_p/C_V$; $C_V,C_P$ being heat capacities per unit
mass under constant volume and pressure correspondingly. Any other
fluid may be considered as well by expansion of thermodynamic
relations into the Taylor series in the vicinity of equilibrium
state \cite{PLK}.

The quasi-plane flow along y-axis will be considered. That allows
to introduce small parameter $\mu$ expressing the relation between
characteristic longitudinal (denoted by $\lambda$) and transverse
scales, for simplicity the same for both transversal directions x
and z. The equivalent system in the dimensionless variables :
\begin{equation}\label{3}
\begin{array}c
   \overrightarrow
{v}_{\ast},\overrightarrow{x}_{\ast},\rho_{\ast},p_{\ast},t_{\ast
}:\overrightarrow{v}= c\overrightarrow{v}_{\ast},p^{\prime
}=c^{2}\rho_{0}p_{\ast},\rho^{\prime}=\rho_{0}%
\rho_{\ast},\\
\overrightarrow{x}=(\lambda
x_{\ast}/\sqrt{\mu},\lambda y_{\ast },\lambda
z_{\ast}/\sqrt{\mu}),t=\lambda t_{\ast}/c,
\end{array}
\end{equation}
(unperturbed values marked by index zero, perturbed ones are
primed, $c=\sqrt{\gamma \frac{p_{0}}{\rho_{0}}}$ is adiabatic
sound velocity), looks as follows (asterisks for dimensionless
variables are omitted here and everywhere later):
\begin{equation}\label{4}
 \frac{\partial}{\partial
t}\psi+L\psi=\varphi+\varphi_{tv},
\end{equation}
where\ $\psi$ is a column of the dimensionless perturbations
\begin{equation}\label{5}
 \psi=\left(\begin{array}
[c]{ccccc}%
v_{x} & v_{y} & v_{z} & p & \rho
\end{array}\right)^{T}.
\end{equation}
$L$ is a linear matrix operator :
\begin{equation}\label{6}
L=\left(
\begin{array}
[c]{ccccc}%
-\delta_{1}^{1}\mu\frac{\partial^{2}}{\partial
x^{2}}-\delta_{1}^{2}\Delta &
-\delta_{1}^{1}\sqrt{\mu}\frac{\partial^{2}}{\partial x\partial y}
& -\delta_{1}^{1}\mu\frac{\partial^{2}}{\partial x\partial z} &
\sqrt{\mu
}\partial/\partial x & 0\\
-\delta_{1}^{1}\sqrt{\mu}\frac{\partial^{2}}{\partial x\partial y}
& -\delta_{1}^{1}\frac{\partial^{2}}{\partial
y^{2}}-\delta_{1}^{2}\Delta &
-\delta_{1}^{1}\sqrt{\mu}\frac{\partial^{2}}{\partial y\partial z}
&
\partial/\partial y & 0\\
-\delta_{1}^{1}\mu\frac{\partial^{2}}{\partial x\partial z} & -\delta_{1}%
^{1}\sqrt{\mu}\frac{\partial^{2}}{\partial z\partial y} & -\delta_{1}^{1}%
\mu\frac{\partial^{2}}{\partial z^{2}}-\delta_{1}^{2}\Delta &
\sqrt{\mu
}\partial/\partial z & 0\\
\sqrt{\mu}\partial/\partial x & \partial/\partial y & \sqrt{\mu}%
\partial/\partial z & -\delta_{2}^{1}\Delta & -\delta_{2}^{2}\Delta\\
\sqrt{\mu}\partial/\partial x & \partial/\partial y & \sqrt{\mu}%
\partial/\partial z & 0 & 0
\end{array}
\right)
\end{equation}
with dimensionless parameters originated by thermal conductivity
and viscosity

$$\delta_{1}^{1}=\frac{\left(  \zeta+\eta/3\right)  }{\rho_{0}c\lambda}%
,\delta_{1}^{2}=\frac{\eta}{\rho_{0}c\lambda},\delta_{2}^{1}=\frac{\chi
}{\rho_{0}c\lambda C_{v}},\delta_{2}^{2}=-\frac{\chi
}{\rho_{0}c\lambda C_{v}\gamma}.$$ There introduced also
dimensionless operators $\vec{\nabla},\Delta$ $\ $:
$\vec{\nabla}=\left(
\begin{array}
[c]{ccc}%
\sqrt{\mu}\partial/\partial x & \partial/\partial y &
\sqrt{\mu}\partial/\partial z
\end{array}
\right)  ,$ $\Delta=\mu\partial^{2}/\partial
x^{2}+\partial^{2}/\partial y^{2}+\mu\partial^{2}/\partial z^{2}.$
The right-hand side of Eq.(\ref{4}) consists of two quadratic
columns, the first $\varphi$ which does not depend on
thermoviscous effects:
\begin{equation}\label{7}
   \varphi=\left(
\begin{array}
[c]{c}%
-(\vec{v}\vec{\nabla})v_{x}+\sqrt{\mu}\rho\partial p/\partial x\\
-(\vec{v}\vec{\nabla})v_{y}+\rho\partial p/\partial y\\
-(\vec{v}\vec{\nabla})v_{z}+\sqrt{\mu}\rho\partial p/\partial z\\
 -\gamma p(\vec{\nabla}\vec{v})-(\vec{v}\vec{\nabla})p\\
-\rho(\vec{\nabla}\vec{v})-(\vec{v}\vec{\nabla})\rho
\end{array}
\right),
\end{equation}
 and the second one $\varphi_{tv}$
appearing in the thermoviscous flow:
\begin{equation}\label{8}
\varphi_{tv}=\left(
\begin{array}
[c]{c}%
-\delta_{1}^{2}\rho\Delta
v_{x}-\delta_{1}^{1}\rho\frac{\partial}{\partial
x}(\vec{\nabla}v)\\
-\delta_{1}^{2}\rho\Delta
v_{y}-\delta_{1}^{1}\rho\frac{\partial}{\partial
y}(\vec{\nabla}v)\\
-\delta_{1}^{2}\rho\Delta
v_{z}-\delta_{1}^{1}\rho\frac{\partial}{\partial
z}(\vec{\nabla}v)\\
(\gamma-1)\left(  \left( \delta_{1}^{1}-\delta_{1}^{2}/3\right)
\left( \vec{\nabla}\vec{v}\right)
^{2}+\frac{\delta_{1}^{2}}{2}\left(
\frac{\partial v_{i}}{\partial x_{k}}+\frac{\partial v_{k}}{\partial x_{i}%
}-\frac{2}{3}\delta_{ik}\frac{\partial v_{l}}{\partial
x_{l}}\right) ^{2}\right)
\end{array}
\right).
\end{equation}
Linear flow is defined by the linearized version of the system of
Eqs(\ref{4})
\begin{equation}\label{9}
  \frac{\partial}{\partial t}\psi+L\psi=0.
\end{equation}
For linear flows, a solution may be found as a sum of planar
waves: $\ v_{x}=\widetilde {v_{x}}(\overrightarrow{k})\exp(i\omega
t-i\overrightarrow{k}\overrightarrow {x}),....$ with wave vector
$\overrightarrow{k}=(k_{x},k_{y},k_{z}).$ In the space of Fourier
transforms (marked by tilde), $-ik_{x}$ means $\partial/\partial
x$, $\left( -ik_{y}\right) ^{-1}$ represents
 $\int dy$, $i\omega$ means
$\partial/\partial t$, etc. System (\ref{9} ) yields in the five
roots of dispersion relation:
\begin{equation}\label{10}
\omega_{1}=\Omega+i\beta\Omega^{2}/2, \omega_{2}=-\Omega+i\beta
\Omega^{2}/2, \omega_{3}=-i\delta_{2}^{2}\Omega^{2}, \omega
_{4}=\omega_{5}=i\delta_{1}^{2}\Omega^{2},
\end{equation}
 where
$\beta=\delta_{1}^{1}+\delta_{1}%
^{2}+\delta_{2}^{1}+\delta_{2}^{2},
  \Omega=k_{y}+\frac{\mu(k_{x}^{2}+k_{z}^{2})}{2k_{y}}.$
These five frequencies relate to three branches of possible
motions in fluid, two acoustic modes, the entropy mode and two
vortical modes. For the real substances,
$\beta>0$,$\delta_{1}^{2}>0$ and $\delta_{2}^{2}<0$, that provides
correct signs of imaginary parts of all frequencies. Modes as
eigenvectors of a linear problem in the space of Fourier
transforms look:

\bigskip$\widetilde{\psi}_{1}=\left(
\begin{array}
[c]{c}%
\widetilde{v_{x}}_{1}(k_{x},k_{y},k_{z})\\
\widetilde{v_{y}}_{1}(k_{x},k_{y},k_{z})\\
\widetilde{v_{z}}_{1}(k_{x},k_{y},k_{z})\\
\widetilde{p}_{1}(k_{x},k_{y},k_{z})\\
\widetilde{\rho}_{1}(k_{x},k_{y},k_{z})
\end{array}
\right)  =\left(
\begin{array}
[c]{c}%
\sqrt{\mu}k_{x}/k_{y}\\
1-\mu(k_{x}^{2}+k_{z}^{2})/(2k_{y}^{2})+i\beta k_{y}/2\\
\sqrt{\mu}k_{z}/k_{y}\\
1+i(\delta_{2}^{1}+\delta_{2}^{2})k_{y}\\
1
\end{array}
\right) \widetilde{\rho}_{1},$
\begin{equation}\label{11}
   \widetilde{\psi}_{2}=\left(
\begin{array}
[c]{c}%
-\sqrt{\mu}k_{x}/k_{y}\\
-1+\mu(k_{x}^{2}+k_{z}^{2})/(2k_{y}^{2})+i\beta k_{y}/2\\
-\sqrt{\mu}k_{z}/k_{y}\\
1-i(\delta_{2}^{1}+\delta_{2}^{2})k_{y}\\
1
\end{array}
\right) \widetilde{\rho}_{2},
\end{equation}

$\widetilde{\psi}_{3}=\left(
\begin{array}
[c]{c}%
0\\
-i\delta_{2}^{2}k_{y}\\
0\\
0\\
1
\end{array}
\right) \widetilde{\rho}_{3},$ $\ \widetilde{\psi}_{4}=\left(
\begin{array}
[c]{c}%
ik_{y}\\
-i\sqrt{\mu}k_{x}\\
0\\
0\\
0
\end{array}
\right)  \widetilde{\sigma}_{4},\ \widetilde{\psi}_{5}=\left(
\begin{array}
[c]{c}%
0\\
-i\sqrt{\mu}k_{z}\\
ik_{y}\\
0\\
0
\end{array}
\right) \widetilde{\sigma}_{5}.$

As basic variable (which all other perturbations are expressed
through) the Fourier transform of density perturbation is chosen
for the first three modes. Vorticites ${\sigma}_{4},{\sigma}_{5}$
are chosen for the last two because the vortical modes possess no
density and pressure perturbations. All calculations of the modes
and projectors have accuracy up to the terms of order $\mu$,
$\beta$. Specific features of the modes follow from Eqs(\ref{11}
): both vorticity modes keep unperturbed density, entropy mode is
isobaric, and so on. Any field of the linear flow is sought as a
sum of independent modes. To calculate projectors that decompose a
concrete mode from the overall field $\widetilde{\psi}$, matrix M
has to be defined:
\begin{equation}\label{12}
\widetilde{M}=(
  \begin{array}{ccccc}
  \widetilde{\rho}_{1} & \widetilde{\rho}_{2} & \widetilde{\rho}_{3} & \widetilde{\sigma}_{4} & \widetilde{\sigma}_{5}) ^T
\end{array}=\widetilde{\psi}=\underset{i=1}{\overset{5}{\sum}}\widetilde{\psi
}_{i},
\end{equation}
 as well as the inverse matrix $\widetilde{M}^{-1}$ .
 Projectors that decompose the overall perturbations into specific
 modes
\begin{equation}\label{13}
    \widetilde{P}_{i}\widetilde{\psi}=\widetilde{\psi}_{i}, i=1,..5
\end{equation}
may be calculated as a product of a column with number $i$ of
$\widetilde{M}$ and row number $i$ of the inverse matrix
$\widetilde{M}^{-1}$:
\begin{equation}\label{14}
   \widetilde{P}_{i}=\widetilde{M}_{i}\cdot\widetilde{{M}^{i}}^{-1},i=1,..5.
\end{equation}

Calculations provided with accuracy of order $\mu,\beta$ yields in
projectors as follows:

{\small $\widetilde{P}_{1,2}= \left(
\begin{array}
[c]{ccccc}%
\mu\frac{k_{x}^{2}}{2k_{y}^{2}} & \sqrt{\mu}\frac{k_{x}}{2k_{y}} &
\mu
\frac{k_{x}k_{z}}{2k_{y}^{2}} & \pm\sqrt{\mu}\frac{k_{x}}{2k_{y}} & 0\\
\sqrt{\mu}\frac{k_{x}}{2k_{y}} & \frac{1}{2}\left(  1\pm\frac{i\beta}{2}%
k_{y}\mp i(\delta_{2}^{1}+\delta_{2}^{2})k_{y}-\mu\frac{k_{x}^{2}+k_{z}^{2}%
}{2k_{y}^{2}}\right)  & \sqrt{\mu}\frac{k_{z}}{2k_{y}} &
\frac{1}{2}\left(
\pm1-i\delta_{2}^{2}k_{y}-\mu\frac{k_{x}^{2}+k_{z}^{2}}{2k_{y}^{2}}\right)
&
\frac{i\delta_{2}^{2}k_{y}}{2}\\
\mu\frac{k_{x}k_{z}}{2k_{y}^{2}} & \sqrt{\mu}\frac{k_{z}}{2k_{y}}
& \mu
\frac{k_{z}^{2}}{2k_{y}^{2}} & \pm\sqrt{\mu}\frac{k_{z}}{2k_{y}} & 0\\
\pm\sqrt{\mu}\frac{k_{x}}{2k_{y}} & \pm\frac{1}{2}\left(  1-\mu\frac{k_{x}%
^{2}+k_{z}^{2}}{2k_{y}^{2}}\right)  &
\pm\sqrt{\mu}\frac{k_{z}}{2k_{y}} &
\frac{1}{2}(1\mp\frac{i\beta}{2}k_{y}\pm i\delta_{2}^{1}k_{y}) &
\pm
\frac{i\delta_{2}^{2}k_{y}}{2}\\
\pm\sqrt{\mu}\frac{k_{x}}{2k_{y}} & \frac{1}{2}\left(  \pm1-i(\delta_{2}%
^{1}+\delta_{2}^{2})k_{y}\mp\mu\frac{k_{x}^{2}+k_{z}^{2}}{2k_{y}^{2}}\right)
& \pm\sqrt{\mu}\frac{k_{z}}{2k_{y}} &
\frac{1}{2}(1\mp\frac{i\beta}{2}k_{y}\mp
i\delta_{2}^{2}k_{y}) & \pm\frac{i\delta_{2}^{2}k_{y}}{2}%
\end{array}
\right)  , $ }

\begin{equation}\label{15}
\widetilde{P}_{3}=\left(
\begin{array}
[c]{ccccc}%
0 & 0 & 0 & 0 & 0\\
0 & 0 & 0 & i\delta_{2}^{2}k_{y} & -i\delta_{2}^{2}k_{y}\\
0 & 0 & 0 & 0 & 0\\
0 & 0 & 0 & 0 & 0\\
0 & i(\delta_{2}^{1}+\delta_{2}^{2})k_{y} & 0 & -1 & 1
\end{array}
\right),
\end{equation}

$\widetilde{P}_{4}=\left(
\begin{array}
[c]{ccccc}%
1-\mu\frac{k_{x}^{2}}{k_{y}^{2}} & -\sqrt{\mu}\frac{k_{x}}{k_{y}}
& -\mu
\frac{k_{x}k_{z}}{k_{y}^{2}} & 0 & 0\\
-\sqrt{\mu}\frac{k_{x}}{k_{y}} & \mu\frac{k_{x}^{2}}{k_{y}^{2}} & 0 & 0 & 0\\
0 & 0 & 0 & 0 & 0\\
0 & 0 & 0 & 0 & 0\\
0 & 0 & 0 & 0 & 0
\end{array}
\right)  ,\widetilde{P}_{5}=\left(
\begin{array}
[c]{ccccc}%
0 & 0 & 0 & 0 & 0\\
0 & \mu\frac{k_{z}^{2}}{k_{y}^{2}} & -\sqrt{\mu}\frac{k_{z}}{k_{y}} & 0 & 0\\
-\mu\frac{k_{x}k_{z}}{k_{y}^{2}} & \sqrt{\mu}\frac{k_{z}}{k_{y}} &
1-\mu
\frac{k_{z}^{2}}{k_{y}^{2}} & 0 & 0\\
0 & 0 & 0 & 0 & 0\\
0 & 0 & 0 & 0 & 0
\end{array}
\right).$

Matrix projectors satisfy common properties of orthogonal
projectors:
 \begin{equation}\label{16}
   \overset{5}{\underset{i=1}{\sum}}\widetilde{P}_{i}=\widetilde{I},\widetilde{P}_{i}\cdot\widetilde{P}_{n}=\widetilde{0}
\hspace*{3mm}if \hspace*{3mm} i\neq
n,\widetilde{P}_{i}\cdot\widetilde{P}_{i}=\widetilde{P}_{i}\hspace*{3mm}
if \hspace*{3mm}i=n,(i,n=1,..5),
\end{equation}
where $\widetilde{I},\widetilde{0}$ are unit and zero matrices.
The inverse transformation of formulae (\ref{15}) to the
$(\overrightarrow{x},t)$ space may be easily undertaken.

Since modes in linear flow are exactly decomposed by projectors,
the dynamic equation for every mode may be originated from the
system (\ref{9}) by acting of the corresponding projector:
$P\left( \frac{\partial }{\partial t}\psi+L\psi\right) =0.$ The
famous dynamic linear equations for acoustic pressure is as
follows:
  \begin{equation}\label{17}
   \frac{\partial p_{1,2}}{\partial
t}\pm\frac{\partial p_{1,2}}{\partial
y}\pm\frac{\mu}{2}\int\Delta_{\perp}
p_{1,2}dy-\frac{\beta}{2}\frac {\partial^{2} p_{1,2}}{\partial
y^{2}}=0.
\end{equation}

Equations for vortical and entropy modes are ordinary equations of
thermal conductivity and are well-known also and may be found in
many sources \cite{BZZ}. Acting by projectors at the original
system of Eqs(\ref{4}) with non-zero nonlinear part (that
essentially depends on all modes) results in coupled equations for
interacting modes: $P\left( \frac{\partial}{\partial
t}\psi+L\psi\right)  = P\left( \varphi_{1}+\varphi_{1tv}\right) $.
When one of the modes is dominant and role of all other ones in
quadratic source is ignored, acting of the corresponding projector
yields in nonlinear dynamic equations, the famous Earnshow one for
the plane flow ($\mu\rightarrow 0$), the
Khokhlov-Zabolotskaya-Kuznetsov (KZK) equation  for acoustic
pressure in quasi-plane viscous flow as follows:
\begin{equation}\label{18}
   \frac{\partial p_{1,2}}{\partial t}\pm\frac{\partial p_{1,2}}{\partial
y}\pm\frac{\mu}{2}\int\Delta_{\perp}
p_{1,2}dy-\frac{\beta}{2}\frac
{\partial^{2} p_{1,2}}{\partial y^{2}}\pm\frac{\gamma+1}{2}%
p_{1,2}\frac{\partial}{\partial y} p_{1,2}=0,
\end{equation}
and other equations.

\bigskip

\section{Streaming cased by any acoustic source. Radiation force of streaming.}

  The streaming as secondary flow induced by losses of acoustic momentum
and its partial nonlinear transfer to the momentum of vortical
mode is best observable when nonlinear effects may store with
time. Most experiments deal with quasi-periodic waves or wave
packets of many periods. That gives rise to the theoretical basis
concerning namely to this kind of source. An area of projecting is
much more extended that streaming since all possible interactions
of modes may be calculated. From the point of view of nonlinear
interactions, streaming is a particular case of dominant acoustic
source inducing the vortical flow. The governing evolution
equation follows when one acts by sum of projectors $P_{4}$ $+$
$P_{5}$ at the system of Eqs(\ref{4}). Only acoustic terms related
to the first mode (rightwards progressive) should be left at the
both quadratic columns . Physically, that corresponds to the
initial stage of nonlinear interaction when the acoustic mode is
dominant and the inverse influence of the streaming may be
ignored. Also, the generation of heat is ignored. A dynamic
equation for transversal component of velocity $V_{x}$ looks
\begin{equation}\label{STR}
  \frac{\partial V_{x}}{\partial t}-\delta_{1}^{2}\frac{\partial
^{2}V_{x}}{\partial y^{2}}=-(\overrightarrow{V}%
\overrightarrow{\nabla})V_{x}+ F_{1x},
\end{equation}
where $F_{1x}$ is a transverse quadratic acoustic source
(radiation force)caused by the rightwards progressive acoustic
mode. Dropping calculations, the result is :

\begin{equation}\label{F}
 F_{1x}=\sqrt{\mu}\beta\left(  \frac{\partial p_{1}}{\partial x}%
\frac{\partial p_{1}}{\partial y}-0.5 p_{1}\frac{\partial^{2} p_{1}%
}{\partial x\partial y}-\frac{\partial}{\partial x}\int
dy(\frac{\partial p_{1}}{\partial y})^{2}\right)  .
\end{equation}
The radiation force appears in the viscous flow only. As expected,
all quadratic nonlinear terms in the acoustic source of order
$\beta^0$ give zero in sum and therefore attenuation is necessary
for acoustic streaming to exist. Acoustic pressure $p_{1}$ should
be a solution of KZK equation. Projectors $P_{4,}P_{5}$ and their
sum yield in the vortical radiation force automatically due to
their origin.

Let us discuss the result to demonstrate the difference between a
periodic and pulsed acoustic sources.

1. In a case of a pulsed ultrasound as in the papers \cite{St1},
\cite{St2} with a carrier frequency $\omega_0$ the solution of the
equation (\ref{STR}) with the right-hand side of Eq.(\ref{F})
contains relatively small (non-growing) part, oscillating with the
frequency $2\omega_0$. It is known, that there is also a term
growing proportional to time, that arises from constant average of
a squared oscillating functions that is evident from the identity
$Sin^2(\omega_0 t)=(1-Cos(2\omega_0 t))/2$. If one take a
wavetrain of a rectangular form, it gives the input proportional
to $ P_0^2\tau/2$; in the case of a triangular pulse one gets
$P_0^2\tau^2/4$ (both per pulse). The second result is obviously
larger if $\tau > 2\pi/\omega_0$. This simple example explains why
the streaming amplitude depends on pulse structure of the source.
Experimental data proving the larger velocity of streaming caused
by pulsed mode are discussed by Starritt, Duck, and Humphrey
\cite{St1}.

 2. It is very important that now the averaged
force $\langle F_{1x}\rangle$ contains the additional terms that
cancel in the case of a periodic sound field. The integration by
parts highlights the point:
\begin{equation}\label{IBP}
\int_y^{y+\Delta y}(p p_{yy}+p_y^2)dy = 0.5(p^2)_y|_y^{y+\Delta
y},
\end{equation}
the right-hand side is obviously zero for a spatially periodic
function with period $\Delta y$ standing for the acoustic
pressure. This shows non-equivalence of the expressions for the
force obtained by averaging procedure by different authors
delivered in different stages of the streaming equation
derivation. In other words if one adds such expression to the
force, the mean value of it survives in a periodic wave case while
for any deviation of periodicity the term contributes.

Radiation force given by Gusev, Rudenko \cite{GR2} in the special
case of strongly attenuated quasi-periodic acoustic beam with weak
diffraction is given by
\begin{equation}\label{beam}
   \Phi_{1x}= \sqrt{\mu} P_{0}^{2}\frac{\partial}{\partial
x}(\theta^{2}/2)\exp(-\beta y)
\end{equation}

rewritten in our variables. It is value averaged over integer
number of sound periods and relates to acoustic pressure in the
rightwards beam as follows:
\begin{equation}\label{21}
   p_{1}(x,y,t)=P_{0}\theta(x)\exp(-\frac{\beta}{2}y)\sin(t-y).
\end{equation}

Calculating based on formula (\ref{F})results in the following:
 \begin{equation}\label{22}
  F_{1x}-\Phi_{1x}=\frac{P_{0}^2 \beta \sqrt{\mu}}{4}\sin(2(t-y))+O(\beta ^{2}).
\end{equation}

Averaging over integral number of period of the sound wave gives
$\left\langle F_{1x}\right\rangle =\Phi_{1x}$ in the leading
order.

\section{Numerical examples and discussions.}

 In the frames of report by Gusev, Rudenko \cite{GR2}, a strongly attenuated non-diffracting acoustic
beam in unbounded volume is considered. The calculations give
values of frequencies above $1 MHz$ in air in the case of radiator
with a radius of order $0.1 m$. That considerably simplifies the
further calculations since a source should be thought as a
solution of the KZK equation and till now does not have an
appropriate mathematical basis. Even the Kholhlov-Zabotskaya (KZ)
equation suitable for flows without attenuation is extremely
difficult for analytical solution: evaluating of flow in the
paraxial region needs many intermediate calculations \cite{HRK}.
In spite of general formula on radiation force given by
Eq.(\ref{F}), any simple solution is desired for illustrations
though the streaming caused by any other wave satisfying the KZK
equation may be treated by this formula. As an appropriate
acoustic source satisfying the limitations of paper by Gusev,
Rudenko \cite{GR2}, let us take a mono-polar\ two-dimensional wave
as follows \cite{RS}:
\begin{equation}\label{SSS}
  p_{1}(x,y,t)=-\sqrt{\frac{2\beta}{\pi}}\exp(-x^{2})\frac{\exp\left(
-\tau^{2}/2\xi\right)  }{\epsilon\sqrt{\xi/\beta}\left(
C-Erf(\tau/\sqrt {2\xi}\right))  },
\end{equation}

where $\xi=\beta y$, \ $\tau=t-y,$ $\epsilon=\frac{\gamma+1}{2}$.
A self-similar solution like (\ref{SSS}) possesses correlated
values of characteristic scale ad amplitude in contrast to
periodic perturbations where effective Mach number and wave length
may be sought independent. Constant C is responsible for the shape
of the curve: large C provides a curve close to the Gauss one. To
calculate a transverse velocity $V_{1x}$, the first, corresponding
force $F_{1x}$ should be calculated, and the second, the momentum
 equation (\ref{STR}) has to be solved.

 The formula of this chapter  apply to the three-dimensional flow and
 are also suitable in the two-dimensional or plane one. Generally,
 two parameters $\mu_{x}$,$\mu_{z}$ instead of one $\mu$ responsible
 for the geometry of flow may be introduced from the very
 beginning. To consider two-dimensional geometry, the flow over z-axis
  supposed to be uniform. To calculate the field of velocity
 accordingly to this last
 equation, let us do some remarks. It is important that the
 hydrodynamic nonlinearity which is necessary in periodic sources
(otherwise the streaming grows infinitely \cite{GR1} \cite{GR2}
\cite{KKKB}) may be not considered in the pulse dynamics, because
of absence in this case the storage of nonlinear effects.
Amplitudes of the generated
 streaming is so small that the quadratic term such as this one
 may be omitted. Without the mentioned term, Eq.(\ref{STR}) is a simple equation of the heat conductivity
 with viscous coefficient cased by shear viscosity $\delta_{1}^{2}$ with an acoustic source in the right-hand side.

 In order to simplify calculations, the viscous term in Eq.(\ref{STR}) will
 be ignored. The same limit relating to low shear viscosity was
 considered in the referred paper by Gusev, Rudenko \cite{GR2}. Note that viscous
 terms grow significantly near the discontinuity of the profile of
 $V_x$, that never happens to a smooth profile caused by pressure
 source given by Eq.(\ref{SSS}). To calculate the temporal field of velocity at a set of
 co-ordinates y,
 the integration of the radiation force $F_{1x}$ has
 been occurred:
 \begin{equation}\label{26}
  V_{x}=\int_{0}^{t} F_{1x}dt.
\end{equation}
 For any co-ordinate y, ignoring of viscous term results in stationary
 velocity after the source passed. The reason is the constant result of integration of the source from
 minus infinity till plus infinity. Indeed, there is an attenuation in the very Eq.(\ref{STR}),
 not only in the attenuated acoustic source, so for the more detail
 calculations it should be involved. The radiation force also needs integration over $y$, the constant
 of integration is chosen to be zero when $y\rightarrow\infty$.
 Transversal derivatives in formula for the radiation force and $V_{x}$ are
 simple when diffraction is ignored:
 \begin{equation}\label{27}
 \frac{\partial p^{2}_{1}(x,y,t)}{\partial x}=-2x\exp(-x^{2}),
\end{equation}
 so all calculations of velocity in this paper refer to the point
 $x=\frac{\sqrt{2}}{2}$ where the radiation force achieves
 maximum. Constants of Eq.(\ref{SSS}) used in calculations are:
$C=2,\epsilon=1.2,\beta=0.1$

Figure 1 shows the temporal development of velocity $V_{x}$ of
streaming divided by $\sqrt{\mu}\beta$ (one may get dimension
values by multiplying by $c$) at different points y:1, 2, 3,.. 10;
$x=\sqrt{2}/2$. The stationary field of
$\frac{V_{x}}{\sqrt{\mu}\beta}$ after passing the source is
presented by Fig. 2. It is evident that velocity of streaming
tends to a constant level with time passing in contrast to the
streaming caused by periodic source when hydrodynamic nonlinearity
ignored.

The next, the streamlines may be plotted due to the calculated
$V_{x}$ and the known relation for the vector field:
$\overrightarrow{\nabla}\overrightarrow{V}=0$. The following
integration with the proper constant should be occurred over y to
get a field of longitudinal velocity $V_{y}$. Finally, the
calculated streamlines at $t=3,t=5$ are shown at the figures
3,4(a) both with the dimensionless pressure of acoustic source
correspondingly (Fig. 3,4(b)). The symmetrical lines of the curves
are $x=0.5$ at the upper half-space and $x=-0.5$ at the lower one
due to the chosen shape of the source given by $exp(-x^{2})$.

 Some calculations of dimensional parameters of flow seems to be
useful under limitations of report by Gusev, Rudenko \cite{GR2}.
Quantitative evaluation relates to air. For example, a
characteristic length of source is $\Lambda\sim 5\cdot 10^{-5} m$,
that gives a dimensionless $\beta=4\cdot 10^{-6} m/ \Lambda\sim
0.1$. A radius of transducer is $R=0.1 m$. Pressure of acoustic
source and therefore streaming depends on small parameter $\beta$;
the second one , $\mu$ appears as a multiplier in the expression
for the radiation force. If hydrodynamic nonlinearity and shear
viscosity ignored, $V_{x}$  is simply proportional to
$\sqrt{\mu}=\Lambda/R$. Accordingly to the accepted values of
$\Lambda$ and $R$, $\sqrt{\mu}=5\cdot 10^{-4}$. Calculations show
that for effective Mach numbers of source about $4\cdot 10^{-2}$
velocity of streaming achieves stationary level $5\cdot
10^{-4}m/sec$ (see figures 1,2, y=3). Meaning a single pulse as
acoustic source large velocities of streaming are hardly expected
while there is no storage of nonlinear transport of acoustic
momentum over many periods. Nevertheless it looks not extremely
small in comparison to the measured values of streaming that vary
from $10^{-3} m/sec$ till $1 m/sec$ \cite{MO}. The sensitivity of
modern technique enabled streaming velocities down to $10^{-4}
m/sec$ \cite{St2}.

\section{Conclusions}

The most important result of the projecting is to get space and
temporal structure of any mode independently of the type of
source. The basic idea of the projecting starts from separating
modes accordingly to their specific properties in the weakly
nonlinear flow. These basic motions (or eigenvectors of the linear
flow) should be defined at first. It may be proceeded in the
algorithmic way for both homogeneous and inhomogeneous backgrounds
( see paper on interacting modes in bubbly liquid \cite{P1}),
media affected by external forces and the real geometry of the
flow. The definition of modes is unique, determined by the
linearized differential conservation equations only. The flow in
bounded volumes or relating to problems that need special
conditions has to be sought as a superposition of specific modes
corresponding to the concrete problem. Any mode is distinguished
by a correspondent projector at any time. Projectors apply to
arbitrary source and type of initial perturbation.

The next, the interaction of modes in weakly nonlinear flow yields
in coupled nonlinear equations for the modes that may be solved
approximately \cite{P1}, \cite{L}. Moreover, the evolution
equations may be corrected up to the higher order nonlinear terms
due to increasing influence of the other generated modes. In the
present paper, rightwards acoustic mode is sought as a dominant
one which gives rise to the vortical one. Therefore, the governing
equation (\ref{STR}) involves a pure quadratic acoustic source.
For more advanced flow, the growing role of the non-dominative
modes due to nonlinear interactions may be accounted as well. The
similar calculations were undertaken while one-dimensional flow
studied \cite{P1}.

Since the method applies to flow with any initial conditions
(including non-acoustic) and does not need quasi-periodic sources,
it is useful for investigation of some special problems like
streaming caused by the non-periodic mono-polar acoustic source.
The possibilities of analytical methods in the study of streaming
are superior over experimental and purely numerical investigations
in the view of complexity of the whole phenomenon. In this paper,
the general formulae on interacting modes are presented, and the
particular case of streaming caused by mono-polar source is
discussed and illustrated by numerical calculations. A special
meaning of the quasi-periodic and particulary pulse ultrasound is
its importance in medicine, in delicate exploring of the
parameters of fluid as well as importance of the secondary modes
following the source, e.g.in the sonochemistry \cite{DDHR} .

\newpage
\textbf{Figure Captures}

FIG. 1. Dimensionless transversal velocity
$\frac{V_{x}}{\sqrt{\mu}\beta}$ via time at different longitudinal
points y:1,2,3,..10 (from the lowest to the upper curve),
$x=\sqrt{2}/2$.

FIG. 2. Stationary dimensionless transversal velocity
$\frac{V_{x}}{\sqrt{\mu}\beta}$  via longitudinal co-ordinate y at
$x=\sqrt{2}/2$.

FIG. 3. a)Streamlines in the plane (x,y), t=3; x-transversal and y
-longitudinal co-ordinates, b)dimensionless pressure of acoustic
source via y at t=3, $x=\sqrt{2}/2$.

FIG. 4. a)Streamlines in the plane (x,y), t=5; x-transversal and y
-longitudinal co-ordinates, b)dimensionless pressure of acoustic
source via y at t=5, $x=\sqrt{2}/2$.

\newpage
\begin{center}
\epsfig{file=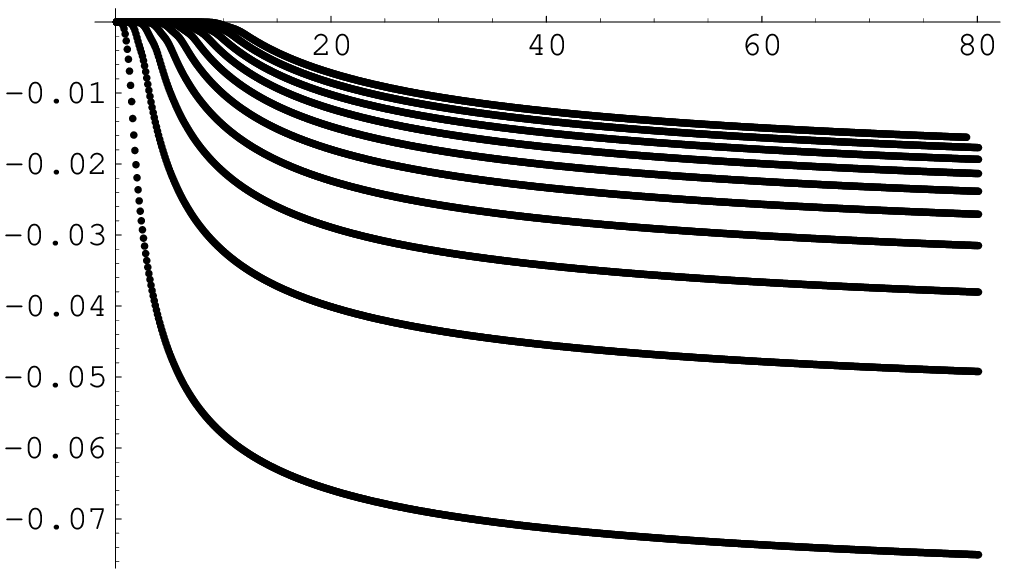, height=4cm, width=8cm ,clip=,angle=0}\\
\end{center}
Perelomova

Figure 1

\newpage
\begin{center}
\epsfig{file=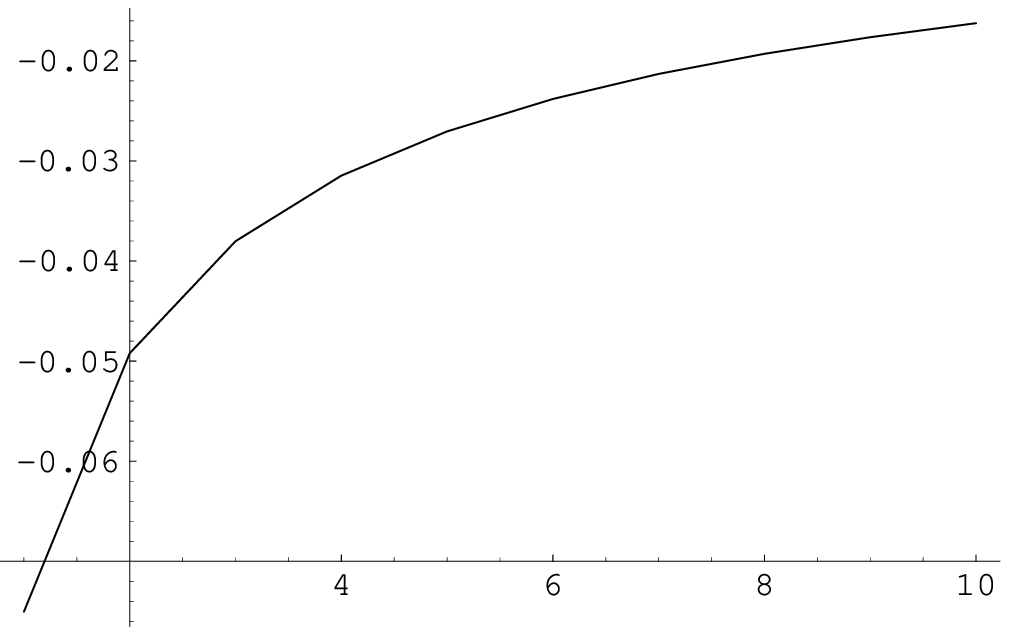, height=4cm, width=6cm ,clip=,angle=0}\\
\end{center}
Perelomova

Figure 2

\newpage
\begin{center}
\epsfig{file=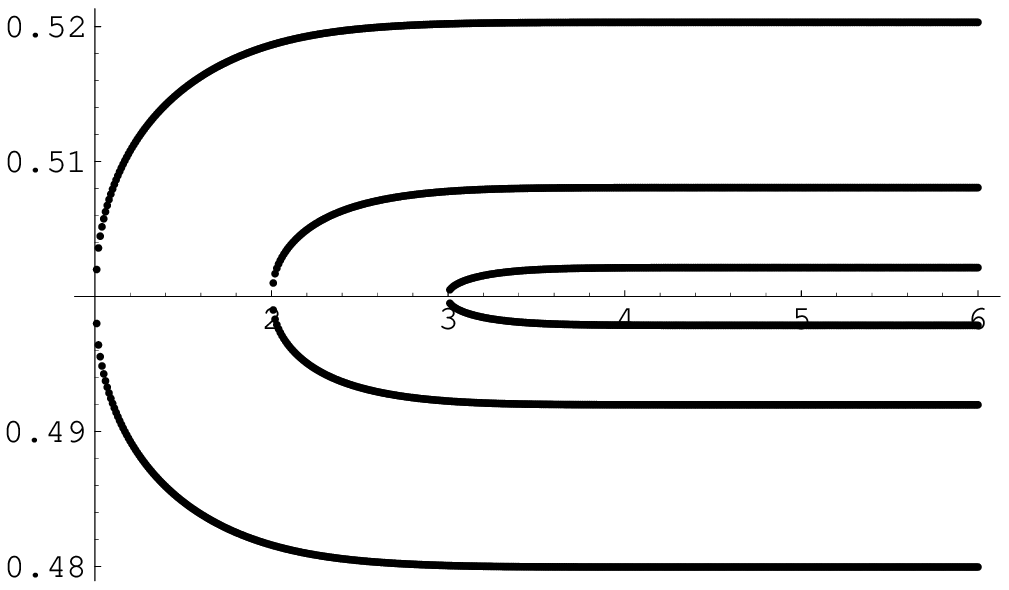, height=4cm, width=6cm,clip=,angle=0}
\hspace*{5mm}
\epsfig{file=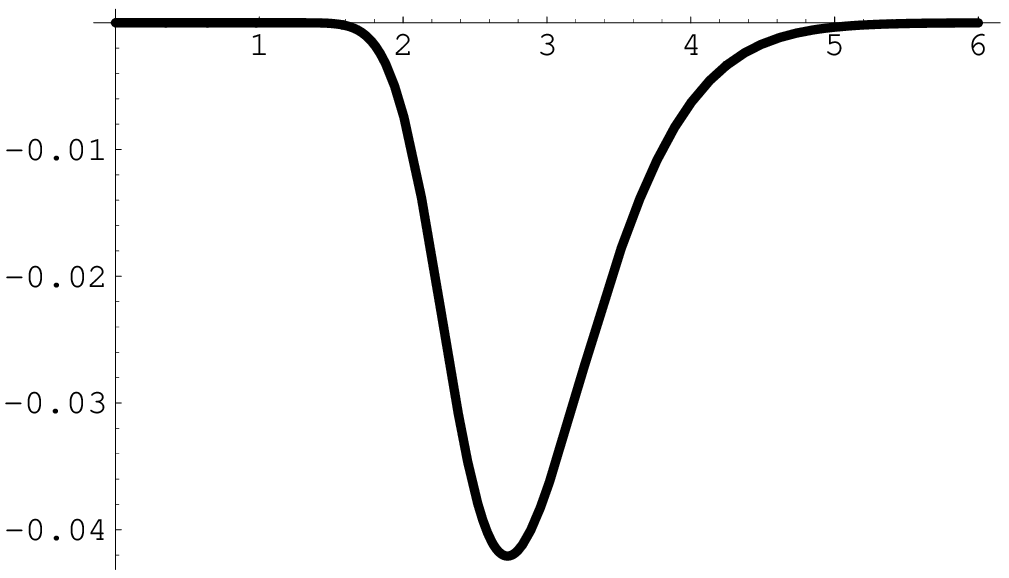, height=4cm, width=6cm ,clip=,angle=0}\\
\end{center}
Perelomova

Figure 3

\newpage
\begin{center}
\epsfig{file=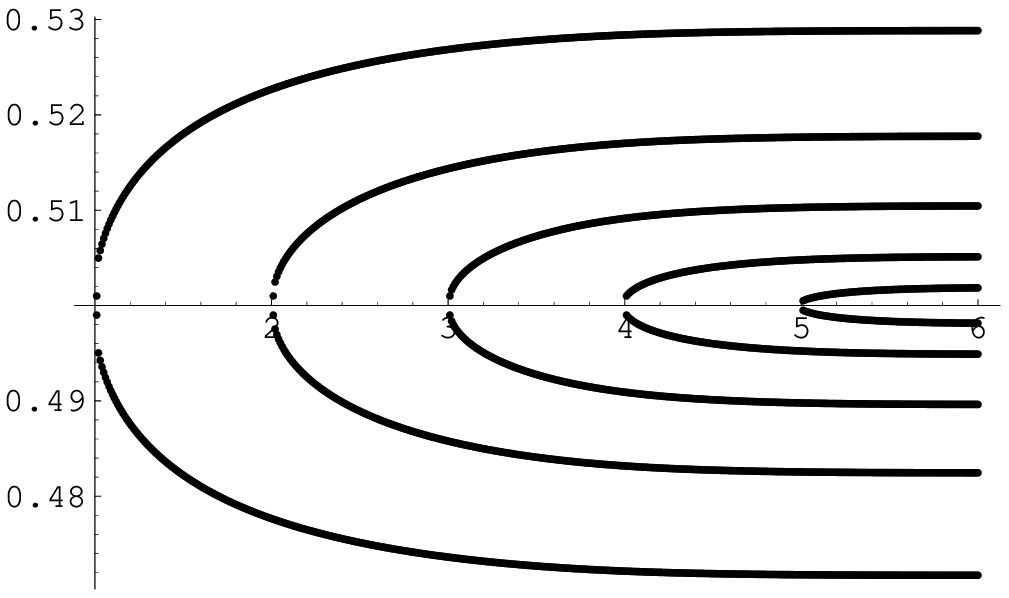, height=4cm, width=6cm ,clip=,angle=0}
\hspace*{5mm}
\epsfig{file=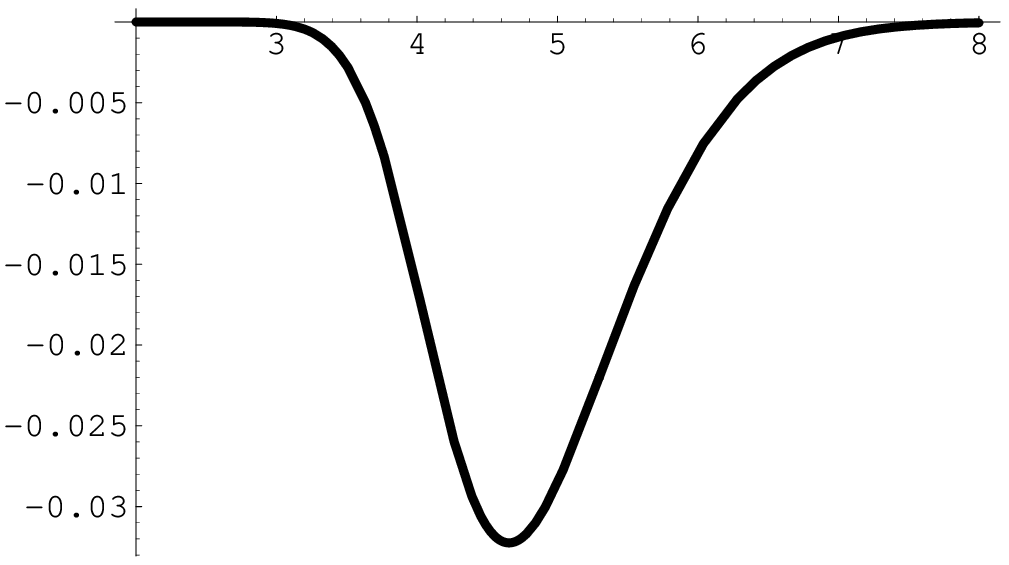, height=4cm, width=6cm ,clip=,angle=0}\\
\end{center}
Perelomova

Figure 4
\end{document}